\documentclass[preprint,12pt]{elsarticle}
\usepackage{amsmath,amsfonts}
\usepackage{algorithmic}
\usepackage{array}
\usepackage{textcomp}
\usepackage{stfloats}
\usepackage{url}
\usepackage{verbatim}
\usepackage{graphicx}
\usepackage{esvect}
\usepackage{algorithm}
\usepackage{algorithmic}
\usepackage{balance}
\usepackage{bm}
\usepackage{multirow}
\usepackage[table,xcdraw]{xcolor}
\usepackage{longtable,booktabs}
\usepackage{multicol,float}
\usepackage{hyperref}
\journal{Signal Processing}

\begin{document}

\begin{frontmatter}

\title{A Comparative Study of Deep Learning and Iterative Algorithms for Joint Channel Estimation and Signal Detection in OFDM Systems\footnote{This work was supported by the National Natural Science Foundation
of China (Grant No. 12101061) and the Young Elite Scientist Sponsorship Program By BAST(No. BYESS2024263).}}

\author[sms]{Haocheng Ju}
\affiliation[sms]{{School of Mathematical Sciences, Peking University, Beijing, 100871, P. R. China}}

\author[bistu]{Haimiao Zhang\corref{cor1}}
 \ead{hmzhang@bistu.edu.cn}

\affiliation[bistu]{{Institute of Applied Mathematics, Beijing
Information Science and Technology University, Beijing, 100192, P. R. China}}

\author[hw]{Lin Li}
\affiliation[hw]{{Huawei Technologies Co., Ltd., Beijing, P. R. China}}

\author[hw]{Xiao Li}

\author[bicmr,cmlr]{Bin Dong\corref{cor1}}
 \ead{dongbin@math.pku.edu.cn}

\affiliation[bicmr]{{Beijing International Center for
Mathematical Research, Peking University, Beijing, 100871, P. R. China}}

\affiliation[cmlr]{{Center for Machine Learning
Research, Peking University, Beijing, 100871, P. R. China}}

\cortext[cor1]{Corresponding authors.}

\begin{abstract}
Joint channel estimation and signal detection (JCESD) is crucial in orthogonal frequency division multiplexing (OFDM) systems, but traditional algorithms perform poorly in low signal-to-noise ratio (SNR) scenarios. Deep learning (DL) methods have been investigated, but concerns regarding computational expense and lack of validation in low-SNR settings remain. Hence, the development of a robust and low-complexity model that can deliver excellent performance across a wide range of SNRs is highly desirable.
In this paper, we aim to establish a benchmark where traditional algorithms and DL methods are validated on different channel models, Doppler, and SNR settings, particularly focusing on the semi-blind setting. In particular, we propose a new DL model where the backbone network is formed by unrolling the iterative algorithm, and the hyperparameters are estimated by hypernetworks. Additionally, we adapt a lightweight DenseNet to the task of JCESD for comparison. We evaluate different methods in three aspects: generalization in terms of bit error rate (BER), robustness, and complexity.
Our results indicate that DL approaches outperform traditional algorithms in the challenging low-SNR setting, while the iterative algorithm performs better in high-SNR settings. Furthermore, the iterative algorithm is more robust in the presence of carrier frequency offset, whereas DL methods excel when signals are corrupted by asymmetric Gaussian noise.\\

\end{abstract}


\end{frontmatter}

\section{Introduction}
Single-input-multiple-output (SIMO) systems play a pivotal role in modern wireless communication standards, including 5G New Radio\cite{basar2019media} and 4G Long Term Evolution\cite{martin2009way}. At the physical layer of the receiver in these systems, channel estimation and signal detection are identified as two critical tasks, both of which can be characterized as inverse problems. Efficient algorithms for these inverse problems are essential for achieving optimal performance in SIMO systems. Traditional signal detection algorithms, such as the maximum likelihood (ML) detector, zero forcing (ZF) detector, linear minimum mean square error (LMMSE) detector, sphere decoder (SD), and the approximate message passing (AMP) algorithms, have been widely used for SIMO systems. However, the detection performance of these algorithms relies heavily on the perfect information of the channel state. Additionally, the traditional algorithms perform well at high-SNR settings, but they often become degraded when the SNR is low \cite{sanzi2003comparative}.

In recent years, deep neural networks (DNNs) have been introduced to solve the inverse problems in wireless communications with the goal of improving the performance of traditional algorithms. These deep models can be broadly categorized into three groups based on their tasks: 1) channel estimation networks \cite{soltani2019deep,li2019deep,balevi2020high,arvinte2022score,He2018DeepLC}; 2) signal detection networks \cite{Samuel2019LearningTD,He2018AMD,Khani2020AdaptiveNS,Tan2018ImprovingMM,Tan2018LOWCOMPLEXITYMP,goutay2020deep}; and 3) joint channel estimation and signal detection (JCESD) networks \cite{ye2017power,Honkala2021DeepRxFC,yi2020deep,he2020model}. The first group of neural networks aims to recover accurate channel states, often outperforming traditional algorithms, but with higher complexity. The second group performs well in signal detection tasks but relies heavily on the perfect channel state information and may suffer from degraded performances due to channel estimation errors. Instead of optimizing individual components, the authors of DeepRx \cite{Honkala2021DeepRxFC} showed that jointly optimizing the channel estimator and signal detector enhances the overall receiver performance. Their proposed approach, which recovers transmitted bits directly from the received data using a residual network, outperforms the LMMSE receiver with higher computational costs for SNR values ranging from $-4$ dB to $32$ dB. We classify this method as purely data-driven since it relies solely on input data to predict the transmitted bits, without any domain knowledge. Another representative JCESD network \cite{he2020model} is an unrolled dynamics (UD) model that involves unrolling iterative algorithms for JCESD and optimizing the hyperparameters based on the given channel model. The UD model demonstrates superior performance compared to the corresponding iterative algorithm, over a range of SNR values from $0$ dB to $35$ dB.

Although existing DL methods for JCESD have demonstrated improved performance over traditional algorithms in high-SNR settings, their performances in low-SNR settings are rarely discussed. Moreover, the major challenges encountered in DL are its robustness\cite{antun2020instabilities,colbrook2022difficulty,genzel2022solving,gottschling2020troublesome} and high computational costs. Therefore, it is desirable to develop a low-complexity and robust model that can perform well over a wide range of SNRs. In this paper, we aim to establish a benchmark for evaluating traditional algorithms, purely data-driven methods, and UD models under the same semi-blind JCESD test setting for OFDM systems\footnote{Code is available at \href{https://github.com/j991222/MIMO_JCESD}{https://github.com/j991222/MIMO\_JCESD.}}. For the purely data-driven methods, we utilize DeepRx\cite{Honkala2021DeepRxFC} and a lightweight DenseNet\cite{huang2017densely} with only $213k$ parameters, which is approximately one-sixth of that in DeepRx. As for the UD models, we unroll the iterative algorithm, which will be described in Subsection \ref{subsec:iterative}, to form the backbone network, and we predict the hyperparameters, including the noise variance, channel correlation coefficient in the frequency and time dimensions, using hypernetworks. We validate these methods on various channel models with SNR ranging from $-10$ dB to $30$ dB, and compare their generalization performances under two settings: in-distribution and out-of-distribution (OOD). In the in-distribution setting, the testing channel model is the same as the training channel model, while in the OOD setting, the testing channel model differs from the training channel model. In addition, we evaluate the robustness of traditional algorithms and DL models on the data with carrier frequency offset and asymmetric Gaussian noise interference.

The contributions of this paper are summarized as follows:

\begin{itemize}
    \item We benchmark the traditional algorithms and DL methods for semi-blind JCESD in OFDM systems across various channel models and SNR ranges. Our findings show that DL methods perform better in the more challenging low-SNR setting, while the iterative algorithm outperforms in the high-SNR setting with significantly lower complexity.
    \item We propose a new UD model Hyper-WienerNet, which adopts hypernetworks to infer the noise variance and channel correlation coefficients in the iterative algorithm. The proposed Hyper-WienerNet outperforms iterative algorithms and purely data-driven methods in OOD settings.
    \item Our experiments highlight the robustness challenges faced by DL methods for semi-blind JCESD. In particular, we observed that the iterative algorithm is considerably more robust to carrier frequency offset than DL methods. Nonetheless, DL methods exhibit superior performance over traditional algorithms in the presence of asymmetric Gaussian noise interference.
\end{itemize}

The remaining part of the paper is organized as follows. In Section \ref{sec:related-work}, we review the related works on channel estimation and signal detection. Section \ref{sec:sysmodel} describes the inverse problem we study in this work. The traditional approaches for channel estimation and signal detection are reviewed in Section \ref{sec:traditional-models}. Section \ref{sec:deep-learning-methods} presents the newly proposed DL models for joint channel estimation and signal detection problem. Numerical results are discussed in Section \ref{sec:experiments}. We conclude this paper in Section \ref{sec:conclusion}.

\section{Related Work}\label{sec:related-work}

\subsection{Channel Estimation}
Most of the channel estimation algorithms are based on the least squares (LS) and LMMSE estimators\cite{tse2005fundamentals,ma2014data,li2017channel}. Noisy LS channel estimates are obtained at pilot positions, and then interpolated to the whole time-frequency grid via different methods, including Wiener filtering\cite{Hoeher1997TwodimensionalPC}, interpolated fast Fourier transform\cite{schoukens1992interpolated} and polynomial interpolation\cite{wang2001ofdm}. These traditional methods are easy to implement, but their performances are unsatisfactory in the low-SNR setting.

Recently, DL methods have been applied to the channel estimation task.  The authors of ChannelNet\cite{soltani2019deep} considered the channel estimate at pilot positions as a low-resolution image and applied image super-resolution and denoising techniques to obtain a complete estimate. Building upon the approach of ChannelNet\cite{soltani2019deep}, ReEsNet \cite{li2019deep} is a more computationally efficient solution formed by replacing the super-resolution and denoising networks with a lightweight residual network. Another popular approach is to use generative models for channel estimation. In \cite{balevi2020high}, a generative adversarial network was pre-trained on simulated channels, and the authors used the generative prior to solve a non-convex optimization problem in the latent space to estimate the channel at the inference stage. In \cite{dong2020channel}, a conditional generative adversarial network was trained to output the estimated channel directly from the received signal. In \cite{arvinte2022mimo} and \cite{arvinte2022score}, a score network was trained to learn the score function of the channel distribution and was then used to simulate the annealed Langevin dynamics to obtain samples from the posterior at test time. Algorithm unrolling is another popular approach for channel estimation. The authors of \cite{He2018DeepLC} introduced a UD model formed by unrolling the denoising-based approximate message passing (D-AMP)\cite{Metzler2014FromDT} algorithm and replacing the denoiser module with DnCNN\cite{Zhang2016BeyondAG}. Later, many other algorithms were unrolled, including matching pursuit\cite{Yassine2020mpNetVD}, the generalized expectation consistent signal recovery (GEC-SR) algorithm\cite{He2020BeamspaceCE}, and the alternating direction method of multipliers (ADMM)\cite{Mao2021JointCE}. These UD models are more interpretable and have good generalization abilities.


\subsection{Signal Detection}
For the signal detection problem, one often assumes the channel state information (CSI) is perfectly known. Therefore, this problem can be formulated as a linear inverse problem. Models for signal detection can be divided into two categories: 1) the common statistical models that can be solved by efficient optimization algorithms; 2) the rapidly evolving deep neural network models. 

The maximum likelihood detection is an optimal signal detection method in theory, but its computational complexity grows exponentially with the number of detected symbols. Representative signal detection techniques include the zero forcing (ZF) detector, LMMSE detector, SD, AMP\cite{Wu2014LowComplexityID}, and expectation propagation (EP)\cite{Cspedes2014ExpectationPD} algorithms. SD is a search algorithm that performs ML detection locally, reducing the size of the search space. However, its complexity still remains exponential. AMP approximates the loopy belief propagation on the factor graph using a Gaussian distribution, while EP approximates the posterior distribution of symbols by fitting a Gaussian distribution. These iterative algorithms perform well with perfect CSI, but their performance deteriorates significantly with imperfect CSI.

Many classical algorithms for signal detection have been unrolled as neural networks, such as DetNet \cite{Samuel2019LearningTD}, OAMP-Net \cite{He2018AMD}, MMNet \cite{Khani2020AdaptiveNS}, DNN-MPD \cite{Tan2018LOWCOMPLEXITYMP}, DNN-dBP \cite{Tan2018ImprovingMM}, and DNN-MS \cite{Tan2018ImprovingMM}. To improve the generalization of the UD model MMNet \cite{Khani2020AdaptiveNS}, the authors of \cite{goutay2020deep} proposed to infer the network parameters using a hypernetwork \cite{ha2017hypernetworks}. The success of hypernetworks in diverse domains, including natural language processing\cite{ha2017hypernetworks}, neural architecture search\cite{brock2018smash}, medical imaging\cite{zhang2020metainv}, and scientific computing\cite{chen2022meta}, suggests that they hold great potential for improving the performance of UD models. Besides these UD models, there are other DL methods based on generative models. The authors of \cite{he2021learning} proposed to adopt the normalizing flow \cite{rezende2015variational} to model the unknown noise distribution and train the signal detector in an unsupervised fashion. In \cite{sun2020generative}, a generative adversarial network was proposed to estimate the channel transition probabilities for signal detection with unknown symbols. DL methods can often achieve lower BER, although their computational costs are much higher than those of iterative algorithms.

\subsection{Joint Channel Estimation and Signal Detection}
When CSI is unknown, we need to estimate the channels and detect 
signals jointly from the received signals and pilot symbols. Traditional methods for JCESD include linear estimators (e.g., LMMSE) and optimization-based methods such as iterative Wiener filtering \cite{sanzi2003comparative}, maximum-a-posterior-based estimator \cite{Liu2014MAPBasedIC}, AMP \cite{sun2018joint}, and sphere manifold optimization \cite{hong2020semi}. However, hyperparameter selection is often done through trial and error as there is no general principle for it.

To improve the performance of traditional JCESD approaches, DL models have been investigated. One approach is to use standard neural network models to parameterize the mapping from input data to detected symbols\cite{ye2017power} or probability of bits\cite{Honkala2021DeepRxFC}. Another approach is to incorporate channel estimation modules into the network. The authors of \cite{yi2020deep} proposed a complete JCESD architecture by combining a DL-based channel estimator with a DL-based signal detector and training the whole network in an end-to-end manner. These black-box models ignore the domain knowledge and cannot generalize across different channel models. To combine domain knowledge with data-driven methods, UD models have been proposed for JCESD. For example, in OAMP-Net2\cite{he2020model}, the LMMSE channel estimator was combined with a UD model to form a lightweight JCESD architecture that generalizes well across different SNRs ($0$ dB$\sim$ $35$ dB). Recently, diffusion models have been used for joint channel estimation and symbol detection. The authors of \cite{zilberstein2023joint} trained a score network and sampled from the joint posterior distribution of the symbols and channels by running the reverse diffusion process. While demonstrating competitive performance against traditional algorithms, diffusion models take longer inference times as they have to simulate the backward process step-by-step.

\section{Joint Channel Estimation and Signal Detection}\label{sec:sysmodel}
\noindent This paper considers the joint channel estimation and signal detection problem for the single-input and multiple-output (SIMO) systems which are widely used in 5G New Radio\cite{basar2019media} and 4G Long Term Evolution\cite{martin2009way}. Consider a SIMO system with a single transmit antenna and $N_r$ receive antennas, the forward model can be written as:
\begin{equation}
    \begin{aligned}
        &\bm{Y}_{f,s}=\bm{H}_{f,s}\bm{X}_{f,s}+\bm{n}_{f,s}\\
        &f=0,...,F-1; s=0,...,S-1
    \end{aligned}
\end{equation}
where
\begin{itemize}
    \item $F$ and $S$ denote the number of subcarriers and symbols, respectively,
    
    \item $\bm{Y}_{f,s}=(\bm{Y}_{f,s,n_r})_{n_r=0}^{N_r-1}\in\mathbb{C}^{N_r}$ is the received signal in the $f$-th subcarrier and $s$-th symbol.
    
    \item $\bm{H}_{f,s}=(\bm{H}_{f,s,n_r})_{n_r=0}^{N_r-1}\in\mathbb{C}^{N_r}$ is the channel vector in the $f$-th subcarrier and $s$-th symbol.
    
    \item $\bm{X}_{f,s}\in D=\{D_{mn}=(1-2m)
\frac{\sqrt{2}}{2}+(1-2n)
\frac{\sqrt{2}}{2}i|m,n\in\{0,1\}\}$ is the transmitted signal. Every $\bm{X}_{f,s}$ can be represented by 2 bits, $\bm{b}_{f,s,0}$ and $\bm{b}_{f,s,1}$:
\begin{equation}
    \bm{b}_{f,s,0}=(1-\text{sign}(\text{Re}(\bm{X}_{f,s})))/2,\quad    \bm{b}_{f,s,1}=(1-\text{sign}(\text{Im}(\bm{X}_{f,s})))/2
\end{equation}
    \item $\bm{n}_{f,s}\in\mathbb{C}^{N_r}$ is the complex valued Gaussian white noise with zero mean and standard deviation $\sigma$, i.e., $\bm{n}_{f,s}\sim \mathcal{CN}(0,\sigma^2 I)$.
\end{itemize}

The inverse problem is to recover $\bm{X}_{f,s} (f=0,...,F-1; s=0,...,S-1)$ and $\bm{H}_{f,s} (f=0,...,F-1; s=0,...,S-1)$ from the received signal $\bm{Y}_{f,s} (f=0,...,F-1; s=0,...,S-1)$ and pilot symbols $\bm{X}_{f,0} (f=0,2,4,\cdots)$. This nonlinear inverse problem, known as joint channel estimation and signal detection, is different from the channel estimation problem where the transmitted signals $\bm{X}_{f,s}$ are known and we only need to estimate the channel. Also, it is distinct from the signal detection problem, which assumes known channel vectors $\bm{H}_{f,s} (f=0,...,F-1; s=0,...,S-1)$ and concentrates on estimating the transmitted signals.

We assume $\bm{H}_{:,:,i}(0\leq i\leq N_r-1)$ are \emph{i.i.d.} random matrices following a complex Gaussian distribution. Denote the vectorized $\bm{H}_{:,:,i}$ as $\vv{\bm{H}_i}\in\mathbb{C}^{FS},\ \vv{\bm{H}_i}\sim\mathcal{CN}(0,\bm{R_{HH}})$. We also assume $\bm{R_{HH}}$ is the Kronecker product of $\bm{R_s}$ and $\bm{R_f}$, \textit{i.e.,}
\begin{equation}
    \bm{R_{HH}}=\bm{R_s}\otimes \bm{R_f}
\end{equation}
where $\bm{R_s}$ and $\bm{R_f}$ are the covariance matrix of $\bm{H}$ at time and frequency dimension, respectively. $\bm{R_s}$ and $\bm{R_f}$ are Toeplitz matrices and can be calculated as follows:
\begin{equation}
    \bm{R_f}=\mbox{Toeplitz}(\bm{c}^\mathrm{H},\bm{c}),\quad \bm{c}\in\mathbb{C}^F, \qquad \bm{R_s}=\mbox{Toeplitz}(\bm{d}^\mathrm{H},\bm{d}),\quad \bm{d}\in\mathbb{C}^S
\end{equation}
where $\bm{c}$ and $\bm{d}$ are the correlation coefficients in the frequency and time dimension, respectively. In real-world scenarios, $\bm{c}$ and $\bm{d}$ are typically unknown and accurate estimation is necessary.

\textbf{Notations:} We use $(\cdot)^H$ for conjugate transpose of $(\cdot)$. For a 3D tensor $\bm{A}\in\mathbb{C}^{F\times S\times N_r}$, $\bm{A}_{::2,0,:}$ stands for the matrix $(\bm{A}_{2f,0,n_r})_{0\leq f<\lfloor F/2\rfloor,0\leq n_r\leq N_r-1}$. For a 2D matrix $\bm{M}\in\mathbb{C}^{F\times F}$, the submatrix $\bm{M}_{:,::2}$ and $\bm{M}_{::2,::2}$ denote the matrix $(\bm{M}_{i,2j})_{0\leq i\leq F-1,0\leq j\leq \lfloor F/2\rfloor-1}$ and $(\bm{M}_{2i,2j})_{0\leq i,j\leq \lfloor F/2\rfloor-1}$, respectively. For a 3D tensor $\bm{A}\in\mathbb{C}^{F\times S\times N_r}$ and a  2D tensor $\bm{B}\in\mathbb{C}^{F\times S}$, we use $\bm{A}./\bm{B}$ to represent the 3D tensor $\bm{C}\in\mathbb{C}^{F\times S\times N_r}$ and $\bm{C}_{f,s,:}=\bm{A}_{f,s,:}/\bm{B}_{f,s}$.

\section{Traditional Channel Estimation and Signal Detection}\label{sec:traditional-models}
In order to introduce the UD model in Subsection \ref{subsec:ud}, we will describe the non-iterative and iterative algorithms for channel estimation and signal detection in this section.

\subsection{Non-iterative Channel Estimation and Signal Detection}\label{subsec:non-iterative}
We start with the non-iterative channel estimation and signal detection approaches. In this work, we consider the single-symbol configuration of demodulation reference signals (DMRS) 
where the DMRS (or pilot) symbols are inserted at the even positions of the first column of transmitted $\bm{X}\in\mathbb{C}^{F\times S}$, i.e., $\bm{X}_{::2,0}$ is known in the receiver.
The LS channel estimate can be obtained by
\begin{equation}
    \widetilde{\bm{H}}_{f,0,:} = \bm{Y}_{f,0,:}/\bm{X}_{f,0}\label{eq:lsinit}
\end{equation}
where the frequency index $f=0,2,4,\cdots,F-2.$ Then, we apply the Wiener filter to interpolate the LS channel estimate for the first column's data symbol. The interpolated channel estimate is given by
\begin{equation}
    \widehat{\bm{H}}_{:,0,:} = (\bm{R_f})_{:,::2}[(\bm{R_f})_{::2,::2}+\sigma^2\bm{I}]^{-1}\widetilde{\bm{H}}_{::2,0,:}\label{eq:lsinterp}
\end{equation}
where $\bm{R_f}\in \mathbb{C}^{F\times F}$ is the channel correlation matrix in the frequency dimension. The whole channel matrix is estimated by extrapolating the first column estimation as
\begin{equation}
    \widehat{\bm{H}}_{:,s,:} = \widehat{\bm{H}}_{:,0,:},\qquad s=1,\cdots,S-1\label{eq:lscopy}
\end{equation}
Then, an MMSE detector is used for signal detection, which is given by
\begin{equation}
    \widetilde{\bm{X}}_{f,s}=(\widehat{\bm{H}}_{f,s,:}^{\mathrm{H}}\widehat{\bm{H}}_{f,s,:}+\sigma^2 \bm{I})^{-1}\widehat{\bm{H}}_{f,s,:}^{\mathrm{H}}\bm{Y}_{f,s,:}.
\end{equation}

In non-iterative channel estimation, the remaining $S-1$ channels are approximated using the first channel estimation,
which will lead to poor performance in fast time-varying environments. To address this, various iterative channel estimators \cite{sanzi2003comparative,Liu2014MAPBasedIC,Liu2014MAPBI,Ma2004EMBasedCE} have been proposed. In the following subsection, we briefly describe a representative iterative scheme for channel estimation and signal detection \cite{sanzi2003comparative}.

\subsection{Iterative Channel Estimation and Signal Detection}\label{subsec:iterative}
Denote the channel estimate at the $j$-th iteration as $\widehat{\bm{H}}^{(j)}$. $\widehat{\bm{H}}^{(0)}$ is initialized as the non-iterative channel estimate described in Subsection \ref{subsec:non-iterative}. In the $j$-th iteration, MMSE signal detection is performed using $\widehat{\bm{H}}^{(j-1)}$, followed by a projection onto the constellation points, \textit{i.e.},
\begin{align}
&\widetilde{\bm{X}}_{f,s}^{(j)}=((\widehat{\bm{H}}_{f,s,:}^{(j-1)})^{\mathrm{H}}\widehat{\bm{H}}_{f,s,:}^{(j-1)}+\sigma^2 \bm{I})^{-1}(\widehat{\bm{H}}_{f,s,:}^{(j-1)})^{\mathrm{H}}\bm{Y}_{f,s,:}\label{eq:mmseiter} \\
& \widehat{\bm{X}}_{f,s}^{(j)} = (\text{sign}(\text{Re}(\widetilde{\bm{X}}_{f,s}^{(j)}))+i \text{sign}(\text{Im}(\widetilde{\bm{X}}_{f,s}^{(j)})))/\sqrt{2},\label{eq:sign}
\end{align}
where $\text{sign}(\cdot)$ represents the sign function.

The LS channel estimate at the $j$-th iteration is given by
\begin{equation}
    \widetilde{\bm{H}}_{f,s,:}^{(j)} = \bm{Y}_{f,s,:}/\widehat{\bm{X}}_{f,s}^{(j)}\label{eq:ls}
\end{equation}
Then, two one-dimensional Wiener filters are applied to the frequency and time dimension to obtain $\widehat{\bm{H}}^{(j)}$ as
\begin{align}
&\widehat{\bm{H}}^{(j-\frac{1}{2})}_{:,:,n_r}=\bm{R_{f}}(\bm{R_{f}}+\sigma^2 \bm{I})^{-1} \widetilde{\bm{H}}^{(j)}_{:,:,n_r}\label{eq:wienerfreq}\\
&\widehat{\bm{H}}^{(j)}_{:,:,n_r} =  \widehat{\bm{H}}^{(j-\frac{1}{2})}_{:,:,n_r}\bm{R_{s}}(\bm{R_{s}}+\sigma^2 \bm{I})^{-1}\label{eq:wienertime}
\end{align}

Algorithm \ref{alg:iterative} summarizes the iterative channel estimation and signal detection method. The iterative algorithm requires the knowledge of the noise variance $\sigma^2$, the correlation coefficients $\bm{c}\in\mathbb{C}^F$ and $\bm{d}\in\mathbb{C}^S$ in the frequency and time dimension, respectively. While there are many noise estimation techniques in the literature, the correlation coefficient estimators are rarely discussed. Existing noise variance estimators include the the method of second and fourth-order moments\cite{benedict1967joint}, the maximum likelihood estimator\cite{Kay1993FundamentalsOS} and the MMSE estimator\cite{pauluzzi2000comparison,Xu2005ANS}. In the numerical experiments, we adopt the iterative algorithm proposed in \cite{savaux2012iterative} to jointly estimate the noise variance and the channel. Starting from an initial $\widehat{\sigma}^{(0)}$, the noise variance and the channel are updated alternatively as 
\begin{align}
\widehat{\bm{H}}_{::2,0,:}^{(j)} &= (\bm{R_f})_{::2,::2}[(\bm{R_f})_{::2,::2}+(\widehat{\sigma}^{(j)})^2\bm{I}]^{-1}\widetilde{\bm{H}}_{::2,0,:}\\
(\widehat{\sigma}^{(j+1)})^2 &=\frac{1}{2F}\sum_{f=1}^{\frac{F}{2}}\left\|\bm{Y}_{2f,0,:}-\widehat{\bm{H}}_{2f,0,:}^{(j)}\bm{X}_{2f,0}\right\|^2.
\end{align}

The initial $\widehat{\sigma}^{(0)}$ is often selected manually for different channel models and the tuning process can be time-consuming. Recently, some works\cite{He2018DeepLC,Samuel2019LearningTD,He2018AMD,he2020model} proposed to unroll the traditional algorithms in wireless communications and set the hyperparameters to be trainable. Instead of learning the optimal parameters from training data, our approach in Subsection \ref{subsec:ud} learns to estimate the parameters adaptively. Specifically, we use hypernetworks to estimate the noise variance, correlation coefficients in the frequency and time dimension from the input data.

\begin{algorithm}
	\caption{Iterative Channel Estimation and Signal Detection}
	\label{alg:iterative}
	\begin{algorithmic}[1]
		\STATE Compute the LS channel estimate (Equation \eqref{eq:lsinit}).
		\STATE Apply the Wiener filter to the first column's data symbol (Equation \eqref{eq:lsinterp}).
		\STATE Obtain $\widehat{\bm{H}}$ by extrapolating the first data symbol to other columns (Equation \eqref{eq:lscopy}).
		\STATE Set $\widehat{\bm{H}}^{(0)}= \widehat{\bm{H}}$.
		\FOR{$j=1$ to $n$}
		\STATE MMSE detector (Equation \eqref{eq:mmseiter}).
		
		\STATE Projection to constellation points (Equation \eqref{eq:sign}).
		\STATE LS channel estimate (Equation \eqref{eq:ls}).
		\STATE Apply the Wiener filter to $\widetilde{\bm{H}}^{(j)}$ in the frequency dimension (Equation \eqref{eq:wienerfreq}).
		\STATE Apply the Wiener filter to $\widehat{\bm{H}}^{(j-\frac{1}{2})}$ in the time dimension (Equation \eqref{eq:wienertime}).
		\ENDFOR
		\STATE Apply the MMSE detector to obtain $\widetilde{\bm{X}}_{f,s}^{(n+1)}$ (Equation \eqref{eq:mmseiter}).
		\RETURN $\widehat{\bm{H}}^{(n)},\widetilde{\bm{X}}^{(n+1)}$
	\end{algorithmic}  
\end{algorithm}

\section{Deep Learning Methods}\label{sec:deep-learning-methods}
In this section, we introduce several DL methods for JCESD. First, we will illustrate how to apply modern deep neural network architectures to JCESD. Then in Subsection \ref{subsec:ud}, we will describe a specifically designed dynamic neural network, named Hyper-WienerNet, to improve the DNN's generalization ability.

For the input of the neural networks based JCESD models described in this section, we adopt a similar setting as in the prior state-of-the-art method DeepRx\cite{Honkala2021DeepRxFC}. Denote $\bm{X}_p\in\mathbb{C}^{F\times S}$ as the pilot symbol matrix where non-pilot positions are filled with zeros. We concatenate $\bm{Y},\bm{X}_p$ and LS channel estimate $\widetilde{\bm{H}}$ along the third dimension, forming a 3D tensor $\bm{Z}_c\in\mathbb{C}^{F\times S\times (2N_r+1)}$. Then, we stack the real and imaginary parts of $\bm{Z}_c$ to form the neural networks' real-valued input tensor $\bm{Z}\in\mathbb{R}^{F\times S\times (4N_r+2)}$.

\subsection{Purely Data-driven Methods}
The universal approximation theorem\cite{hornik1989multilayer,hornik1991approximation,lu2017expressive} gives the theoretical foundations for the approximation capabilities of fully connected networks (FCN) and convolutional neural networks (CNN). However, it does not provide guidance on choosing an appropriate network architecture or optimizing its parameters to achieve good generalization performance on real-world datasets. Therefore, researchers often rely on empirical methods such as trial and error or heuristics to design neural network architectures and optimization algorithms. In our study, we aim to approximate the mapping from the inputs $\bm{Z}$ to the detected symbols $\bm{X}$ using neural networks, and we will investigate which neural network architectures are suitable for this problem.
These deep models are classified as \emph{purely data-driven methods} since they are designed to predict signals solely from input data without relying on any domain knowledge. In contrast to the traditional non-iterative and iterative algorithms in Subsection \ref{sec:traditional-models}, deep neural network models investigated in this subsection do not need to estimate the
SNR, Doppler, and correlation coefficients.



DenseNet \cite{huang2017densely} is a type of convolutional neural network that utilizes dense connections between layers. It has been shown to achieve comparable performance to ResNet while requiring significantly fewer parameters for the image classification task. To reduce the model complexity of the ResNet used in DeepRx \cite{Honkala2021DeepRxFC}, we employ a DenseNet to predict the probability $P(\Tilde{\bm{b}}_{f,s,d}=1)$ for every bit $\Tilde{\bm{b}}_{f,s,d},\  f=1,\cdots,F,\ s=1,\cdots,S,\ d=0,1$. Therefore, the shape of the DenseNet's output is $F\times S\times 2$. Table \ref{tab:densenet} shows the architecture of DenseNet adopted in our numerical experiment section for the JCESD problem. Note that the kernel size, dilation parameters and the output shape of 2D convolutional neural network layers are listed in the second to fourth columns of Table \ref{tab:densenet}.  For the training of DenseNet, we use the binary cross entropy loss
\begin{equation}
   \mathcal{L}(\Tilde{\bm{b}},\bm{b})=-\frac{1}{2FS}\sum_{f=0}^{F-1}\sum_{s=0}^{S-1}\sum_{d=0}^1  \bm{b}_{f,s,d}\ln P(\Tilde{\bm{b}}_{f,s,d}=1)+
    (1-\bm{b}_{f,s,d})\ln P(\Tilde{\bm{b}}_{f,s,d}=0) 
\end{equation}
where $\bm{b}$ is the ground truth bit, $\Tilde{\bm{b}}$ is the neural network's output.

\begin{table}[htb]
\renewcommand{\arraystretch}{0.7}
\centering
\footnotesize
\begin{tabular}{c|c|c|c|c}
\toprule
Layer Index & Layer  Architecture         & Filter   & Dilation      & Output Shape \\ \midrule
1 &Conv2d           & (3, 3)  &  (1, 1)    & (F, S, 24)     \\ \midrule
2 &Residual Block \uppercase\expandafter{\romannumeral1}  & (3, 3)  &   (1, 1)   & (F, S, 48)     \\ \midrule
3 &Residual Block \uppercase\expandafter{\romannumeral1}  & (3, 3)  & (2, 3)    & (F, S, 96)     \\ \midrule
4 &Residual Block \uppercase\expandafter{\romannumeral2} &  (3, 3) &  (3, 6)  & (F, S, 24)     \\ \midrule
5 &Residual Block \uppercase\expandafter{\romannumeral1} & (3, 3)  &  (1, 1)  & (F, S, 48)     \\ \midrule
6 &Residual Block \uppercase\expandafter{\romannumeral1} & (3, 3)  &  (2, 3)  & (F, S, 96)     \\ \midrule
7 &Residual Block \uppercase\expandafter{\romannumeral2} & (3, 3)  &  (3, 6)  & (F, S, 48)     \\ \midrule
8 &Residual Block \uppercase\expandafter{\romannumeral1}  & (3, 3)  &   (3, 6)   & (F, S, 96)     \\ \midrule
9 &Residual Block \uppercase\expandafter{\romannumeral1}  & (3, 3)  & (2, 3)    & (F, S, 192)     \\ \midrule
10 &Residual Block \uppercase\expandafter{\romannumeral2} &  (3, 3) &  (1, 1)  & (F, S, 48)     \\ \midrule
11 &Residual Block \uppercase\expandafter{\romannumeral1} & (3, 3)  &  (3, 6)  & (F, S, 96)     \\ \midrule
12 &Residual Block \uppercase\expandafter{\romannumeral1} & (3, 3)  &  (2, 3)  & (F, S, 192)     \\ \midrule
13 &Residual Block \uppercase\expandafter{\romannumeral2} & (3, 3)  &  (1, 1)  & (F, S, 48)     \\ \midrule
14 &Conv2d         & (1, 1)  &  (1, 1)  & (F, S, 2)     \\ 
\bottomrule
\end{tabular}
\caption{Network structure of DenseNet. Residual Block \uppercase\expandafter{\romannumeral1} and \uppercase\expandafter{\romannumeral2} are shown in Fig. \ref{fig:densenet}.}
\label{tab:densenet}
\end{table}

\begin{figure}[!t]
\centering
\includegraphics[width=2.5in]{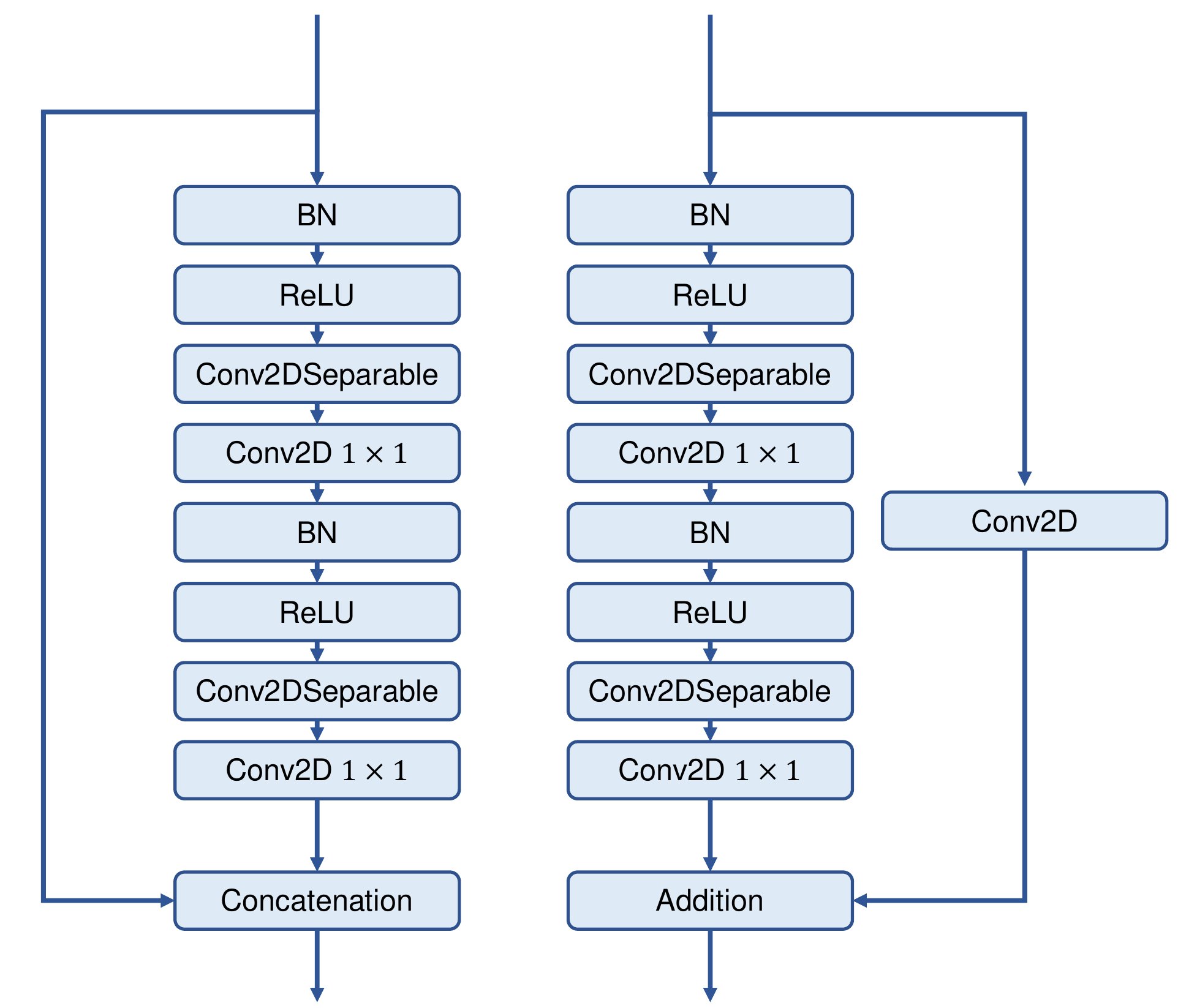}
\caption{Left: Residual Block \uppercase\expandafter{\romannumeral1}; Right: Residual Block \uppercase\expandafter{\romannumeral2}. BN denotes Batch Normalization, and 'Conv2DSeparable' stands for Depthwise Separable Convolution, i.e., each channel uses different kernels to perform convolution.}
\label{fig:densenet}
\end{figure}

\subsection{Hyper-WienerNet}\label{subsec:ud}
\subsubsection{Backbone network}
The backbone network of Hyper-WienerNet is built by unrolling Algorithm \ref{alg:iterative} with $L=6$ iterations. Every time we compute the LS channel estimate or apply the Wiener filter, the noise variance differs from the previous estimate. Therefore, we use different notations for the noise variance in \eqref{eq:ls}, \eqref{eq:wienerfreq} and \eqref{eq:wienertime}. Denote

\begin{align}
    &W_{\gamma_1}=(\bm{R_f})_{:,::2}\left[(\bm{R_f})_{::2,::2}+\gamma_1^2\bm{I}\right]^{-1}\\
    &W_{\gamma_i}=\bm{R_f}(\bm{R_f}+\gamma_i^2\bm{I})^{-1},2\leq i\leq L\\
    &W_{\rho_i}=(\bm{R_s}+\rho_i^2\bm{I})^{-1}\bm{R_s},1\leq i\leq L-1
\end{align}

\begin{figure*}[!t]
\centering
\includegraphics[width=4in]{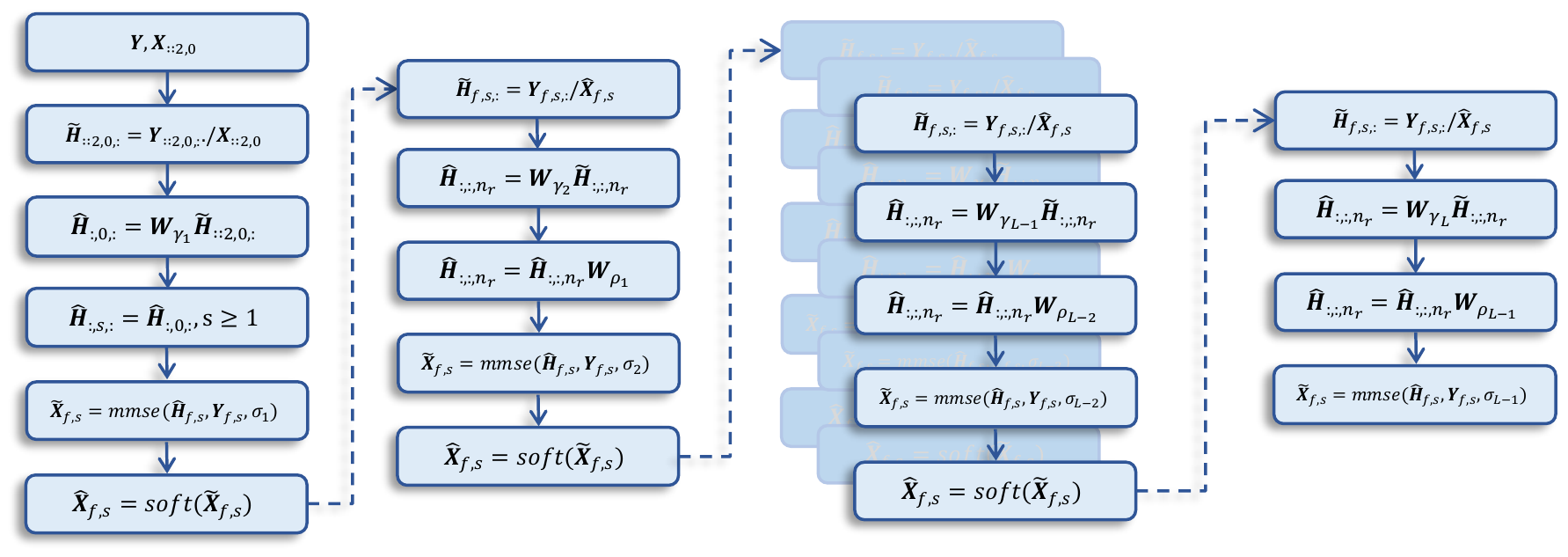}
\caption{Backbone network of Hyper-WienerNet.}
\label{fig:backbone}
\end{figure*}

Fig. \ref{fig:backbone} presents the structure of the backbone network. Here, the module $\text{mmse}(\widehat{\bm{H}}_{f,s},\bm{Y}_{f,s},\sigma_i)$ represents the MMSE detector which is defined by
\begin{equation}
    \widetilde{\bm{X}}_{f,s}=(\widehat{\bm{H}}_{f,s,:}^{\mathrm{H}}\widehat{\bm{H}}_{f,s,:}+\sigma_i^2 \bm{I})^{-1}\widehat{\bm{H}}_{f,s,:}^{\mathrm{H}}\bm{Y}_{f,s,:}.
\end{equation}
The module $\text{soft}(\widetilde{\bm{X}}_{f,s})$ denotes the soft decision of $\widetilde{\bm{X}}_{f,s}$ with the definition as following
\begin{equation}
    \text{soft}(\widetilde{\bm{X}}_{f,s}) = (\text{tanh}(10\text{Re}(\widetilde{\bm{X}}_{f,s}))+i\text{tanh}(10\text{Im}(\widetilde{\bm{X}}_{f,s})))/\sqrt{2}.
\end{equation}
This module is designed to approximate the original sign function sign$(\cdot)$ in \eqref{eq:sign} by a  differentiable function $\text{tanh}(10\cdot)$.

\subsubsection{Hypernetwork}
Since the standard deviation $\gamma_i,\rho_i,\sigma_i$, correlation coefficients $\bm{c}$ and $\bm{d}$ in the frequency and time dimension are unknown in practical applications, we use three hypernetworks $\mathcal{H}_1(\bm{Z};\Theta_1)$, $\mathcal{H}_2(\bm{Z};\Theta_2)$, $\mathcal{H}_3(\bm{Z};\Theta_3)$ to infer these three types of parameters separately, i.e.,

\begin{align}
    (\gamma_1, \sigma_1, \gamma_2, \rho_1,\cdots, \sigma_i, \gamma_{i+1}, \rho_{i},\cdots, \sigma_L )&=\mathcal{H}_1(\bm{Z};\Theta_1)\\
    \bm{c}&=\mathcal{H}_2(\bm{Z};\Theta_2)\\
    \bm{d}&=\mathcal{H}_3(\bm{Z};\Theta_3)
\end{align}

\begin{figure}[!t]
\centering
\includegraphics[width=2.5in]{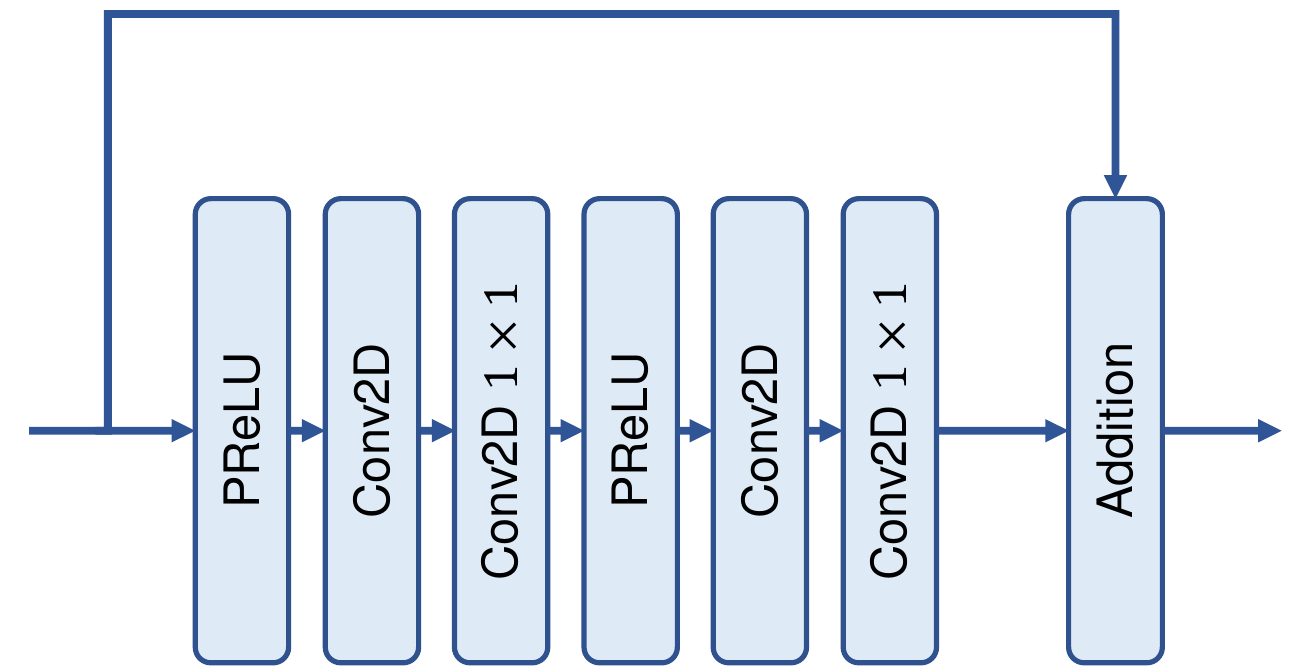}
\caption{Residual Block in $HN_2$.}
\label{fig:resblock}
\end{figure}

\begin{figure*}[!t]
\centering
\includegraphics[width=4in]{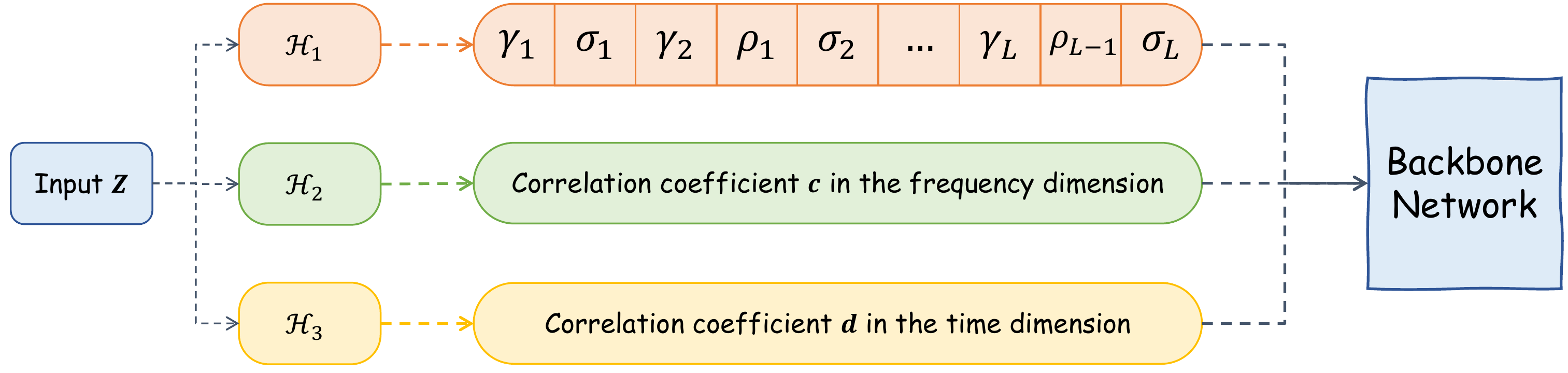}
\caption{Hyper-WienerNet.}
\label{fig:hyperwiener}
\end{figure*}

This approach allows our model to adapt to different data distributions by inferring the unknown values from the received signals and pilot signals. The module $\mathcal{H}_1(\bm{Z};\Theta_1)$ comprises a 2D convolutional layer (Conv2d), Parameterized ReLU layer (PReLU) \cite{he2015delving}, and Linear transform layer (Linear). These neural network layers are arranged orderly as Conv2d $\to$ PReLU $\to$ Linear. The structure of module $\mathcal{H}_2(\bm{Z},\Theta_2)$ is illustrated in Table \ref{tab:hypernetf}. The structure of ``Residual Block" listed in Table \ref{tab:hypernetf} is described in Fig. \ref{fig:resblock}. Here, the Conv2D $1\times 1$ layer means the 2D convolution operation with $1\times 1$ kernel. The module $\mathcal{H}_3(\bm{Z},\Theta_3)$ shares the same structure as $\mathcal{H}_2(\bm{Z},\Theta_2)$ except that the last layer is replaced by averaging over the frequency dimension. The complete hyper-parameter prediction module is shown in Fig. \ref{fig:hyperwiener} and we denote it as Hyper-WienerNet.

\begin{table}[htb]
\renewcommand{\arraystretch}{0.7}
\centering
\footnotesize
\begin{tabular}{c|c|c|c}
\toprule
Layer               & Filter & Dilation & Output Shape \\ \midrule
Conv2d           & (3, 3)  & (1, 1)    & (F, S, 64)     \\ \midrule
Residual Block   &(3, 3)  & (1, 1)   & (F, S, 64)     \\ \midrule
Residual Block  & (3, 3)  & (2, 3)    & (F, S, 128)    \\ \midrule
Residual Block  & (3, 3)  & (3, 6)    & (F, S, 256)    \\ \midrule
Residual Block   & (3, 3)  & (2, 3)    & (F, S, 128)    \\ \midrule
Residual Block  & (3, 3)  & (1, 1)    & (F, S, 64)     \\ \midrule
Residual Block   & (3, 3)  & (1, 1)    & (F, S, 64)     \\ \midrule
Conv2d            & (1, 1)  & (1, 1)    & (F, S, 2)      \\ \midrule
Sigmoid                   &        &          &  (F, S, 2)    \\ \midrule
Average over time dimension              &        &          & (F, 2)        \\
\bottomrule
\end{tabular}
\caption{Network structure of $\mathcal{H}_2$.
}
\label{tab:hypernetf}
\end{table}

\subsubsection{Loss Function}
The newly proposed Hyper-WienerNet's output complex valued signal $\widetilde{\bm{X}}_{f,s}$ is transformed to two bits $\Tilde{\bm{b}}_{f,s,0}$, $\Tilde{\bm{b}}_{f,s,1}$ as follows
\begin{equation}
    \Tilde{\bm{b}}_{f,s,0}=(1-\text{sign}(\text{Re}(\widetilde{\bm{X}}_{f,s})))/2,\qquad \Tilde{\bm{b}}_{f,s,1}=(1-\text{sign}(\text{Im}(\widetilde{\bm{X}}_{f,s})))/2.
\end{equation}

For the neural network training, we adopt the following binary cross entropy loss
\begin{equation}
    \mathcal{L}(\Tilde{\bm{b}},\bm{b},\widetilde{\bm{X}})=-\frac{1}{2FS}\sum_{f=0}^{F-1}\sum_{s=0}^{S-1}\sum_{d=0}^1  \bm{b}_{f,s,d}\ln P(\Tilde{\bm{b}}_{f,s,d}=1|\widetilde{\bm{X}}_{f,s})
    +(1-\bm{b}_{f,s,d})\ln P(\Tilde{\bm{b}}_{f,s,d}=0|\widetilde{\bm{X}}_{f,s})
\end{equation}
where $\bm{b}_{f,s,d}$ is the ground truth bit. The probability of bits $P(\Tilde{\bm{b}}_{f,s,d}=1|\widetilde{\bm{X}}_{f,s})$ and $P(\Tilde{\bm{b}}_{f,s,d}=0|\widetilde{\bm{X}}_{f,s})$ can be computed using the log likelihood ratio (LLR), \textit{i.e.,}
\begin{equation}
P(\Tilde{\bm{b}}_{f,s,d}=1|\widetilde{\bm{X}}_{f,s})=\frac{e^{LLR(\Tilde{\bm{b}}_{f,s,d})}}{1+e^{LLR(\Tilde{\bm{b}}_{f,s,d})}},\qquad P(\Tilde{\bm{b}}_{f,s,d}=0|\widetilde{\bm{X}}_{f,s})=\frac{1}{1+e^{LLR(\Tilde{\bm{b}}_{f,s,d})}}.
\end{equation}

In the following, we will briefly illustrate how to compute the LLR.
Assume the MMSE detector predicts a signal
\begin{equation}
    \widetilde{\bm{X}}_{f,s}=(\widehat{\bm{H}}_{f,s,:}^{\mathrm{H}}\widehat{\bm{H}}_{f,s,:}+\widehat{\sigma}^2 \bm{I})^{-1}\widehat{\bm{H}}_{f,s,:}^{\mathrm{H}}\bm{Y}_{f,s,:}
\end{equation}
Then, we model $ \widetilde{\bm{X}}_{f,s}$ as follows
\begin{equation}
    \widetilde{\bm{X}}_{f,s}=\bm{G}_{f,s} \bm{X}_{f,s}+\bm{n}'_{f,s}
\end{equation}
where $\bm{n}'_{f,s}\sim \mathcal{CN}(0,\bm{\epsilon}_{f,s}^2)$ is the error term, $\bm{\epsilon}_{f,s}^2=\bm{G}_{f,s} (1-\bm{G}_{f,s})$ represents the variance, and $\bm{G}_{f,s}$ is defined by
\begin{equation}
    \bm{G}_{f,s} =(\widehat{\bm{H}}_{f,s,:}^{\mathrm{H}}\widehat{\bm{H}}_{f,s,:}+\widehat{\sigma}^2 \bm{I})^{-1}\widehat{\bm{H}}_{f,s,:}^{\mathrm{H}}\widehat{\bm{H}}_{f,s,:}
\end{equation}
The LLR is calculated as
\begin{align}
    LLR(\Tilde{\bm{b}}_{f,s,d})&=\ln \frac{\sum_{\Tilde{\bm{b}}_{f,s,d}=1}P(\bm{X}_{f,s}|\widetilde{\bm{X}}_{f,s})}{\sum_{\Tilde{\bm{b}}_{f,s,d}=0}P(\bm{X}_{f,s}|\widetilde{\bm{X}}_{f,s})}\\
    &\approx \ln\frac{\max_{\Tilde{\bm{b}}_{f,s,d}=1}P(\bm{X}_{f,s}|\widetilde{\bm{X}}_{f,s})}{\max_{\Tilde{\bm{b}}_{f,s,d}=0}P(\bm{X}_{f,s}|\widetilde{\bm{X}}_{f,s})}\\
    &=-\min_{\Tilde{\bm{b}}_{f,s,d}=1}\frac{\left\|\widetilde{\bm{X}}_{f,s}-\bm{G}_{f,s}  \bm{X}_{f,s}\right\|^2}{\bm{\epsilon}_{f,s}^2}
    +\min_{\Tilde{\bm{b}}_{f,s,d}=0}\frac{\left\|\widetilde{\bm{X}}_{f,s}-\bm{G}_{f,s}  \bm{X}_{f,s}\right\|^2}{\bm{\epsilon}_{f,s}^2},
\end{align}
It is straightforward to show that
\begin{equation}
    LLR(\Tilde{\bm{b}}_{f,s,0})=-\frac{2\sqrt{2}\bm{G}_{f,s}  \text{Re}(\widetilde{\bm{X}}_{f,s})}{\bm{\epsilon}_{f,s}^2},\qquad LLR(\Tilde{\bm{b}}_{f,s,1})=-\frac{2\sqrt{2}\bm{G}_{f,s}  \text{Im}(\widetilde{\bm{X}}_{f,s})}{\bm{\epsilon}_{f,s}^2}.
\end{equation}

\section{Experiments}\label{sec:experiments}
In this section, we compare the performance of traditional algorithms and DL methods on different channel models, Dopplers and SNRs. First, we describe the generation of datasets and the implementation details. Then, we validate the generalization performances of the following methods:
    (1)Non-iterative algorithm described in Subsection \ref{subsec:non-iterative},
    (2)The symbol-by-symbol iterative algorithm in \cite{sanzi2003comparative},
    (3)DeepRx\cite{Honkala2021DeepRxFC},
    (4)DenseNet,
    (5)Hyper-WienerNet.

We investigate two different generalization scenarios: in-distribution and out-of-distribution (OOD). In the in-distribution setting, the testing channel model is the same as the training channel model, while in the OOD setting, the testing channel model differs from the training channel model. Additionally, we assess the robustness of traditional algorithms and DL models to carrier frequency offset and asymmetric Gaussian noise interference. The computational complexity of different methods is also discussed in the end of this section.

\subsection{Datasets}\label{subsec:datasets}
We generate the data using the physical downlink shared channel (PDSCH) in MATLAB's 5G Toolbox. The parameter settings are listed in Table \ref{tab:paramter}. For every triplet (Channel Model, Doppler, SNR) in Table \ref{tab:datasets}, we generate a dataset containing the received signal $\bm{Y}_{f,s}$, transmitted signal $\bm{X}_{f,0}(f=0,2,4,\cdots)$, ground truth channel $\bm{H}_{f,s}$, and ground truth transmitted signal $\bm{X}_{f,s}$. $3,000$ slots of data are generated for each EVA dataset and $600$ slots for the remaining ones.
$60\%$ of the data is used for training, $20\%$ for validation and $20\%$ for testing.

\begin{table}[h]
\renewcommand{\arraystretch}{0.7}
\centering
\footnotesize
\begin{tabular}{c|c}
\toprule Parameter & Value     \\
\midrule Subcarrier Spacing & 30kHz  \\
\midrule Number of Resource Blocks & 51\\
\midrule Modulation Scheme & QPSK\\
\midrule Number of Transmit Antennas & 1\\
\midrule Number of Receive Antennas & 4\\
\midrule Number of Subcarriers & 24\\
\midrule Number of Symbols & 12\\
\bottomrule
\end{tabular}
\caption{Parameter Settings.}
\label{tab:paramter}
\end{table}

\begin{table}[h]
\renewcommand{\arraystretch}{0.7}
\centering
\footnotesize
\begin{tabular}{c|c|c}
\toprule Channel Model & Doppler(Hz) & SNR(dB)    \\
\midrule EVA & 5, 30, 60, 90, 120, 150 & $-10, -5, 0, 10, 20, 30$  \\
\midrule CDL-A & 15, 45, 75, 105, 135, 165 & -8\\
\midrule CDL-E & 15, 45, 75, 105, 135, 165  & -3  \\
\midrule TDL-A & 15, 45, 75, 105, 135, 165  & -8  \\
\midrule TDL-E & 15, 45, 75, 105, 135, 165  & -3  \\
\bottomrule
\end{tabular}
\caption{Datasets.}
\label{tab:datasets}
\end{table}

\subsection{Implementation}
We use the AdamW optimizer with the cosine annealing schedule \cite{loshchilov2017sgdr} to train the aforementioned neural networks. The training settings of the compared models are shown in Table \ref{tab:training}.

\begin{table}[h]
\renewcommand{\arraystretch}{0.7}
\centering
\footnotesize
\begin{tabular}{c|c|c|c}
\toprule             & DeepRx\cite{Honkala2021DeepRxFC} & DenseNet  &Hyper-WienerNet   \\
\midrule Batch Size  &  $300$ &       $300$       &    $1024$         \\
\midrule Learning Rate & $1\times 10^{-3}$  &      $1\times 10^{-3}$     &    $1\times 10^{-4}$          \\
\bottomrule
\end{tabular}
\caption{Training settings for compared neural networks.}
\label{tab:training}
\end{table}

\subsection{Quantitative Results}\label{subsec:quantitative}
\begin{figure*}[!t]
\centering
\includegraphics[width=5in]{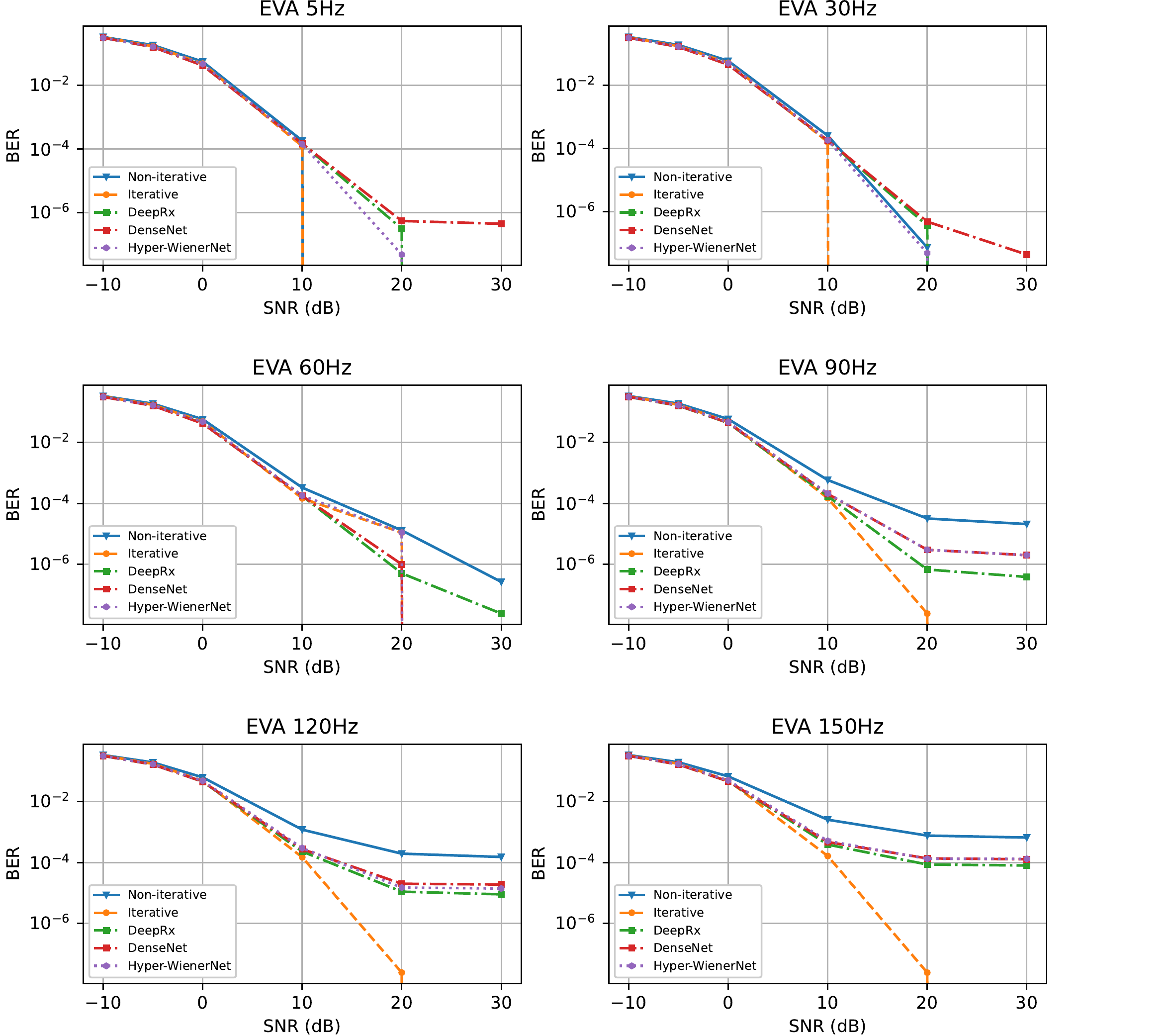}
\caption{In-distribution generalization results.}
\label{fig:eva}
\end{figure*}

\begin{figure*}[!t]
\centering
\includegraphics[width=5in]{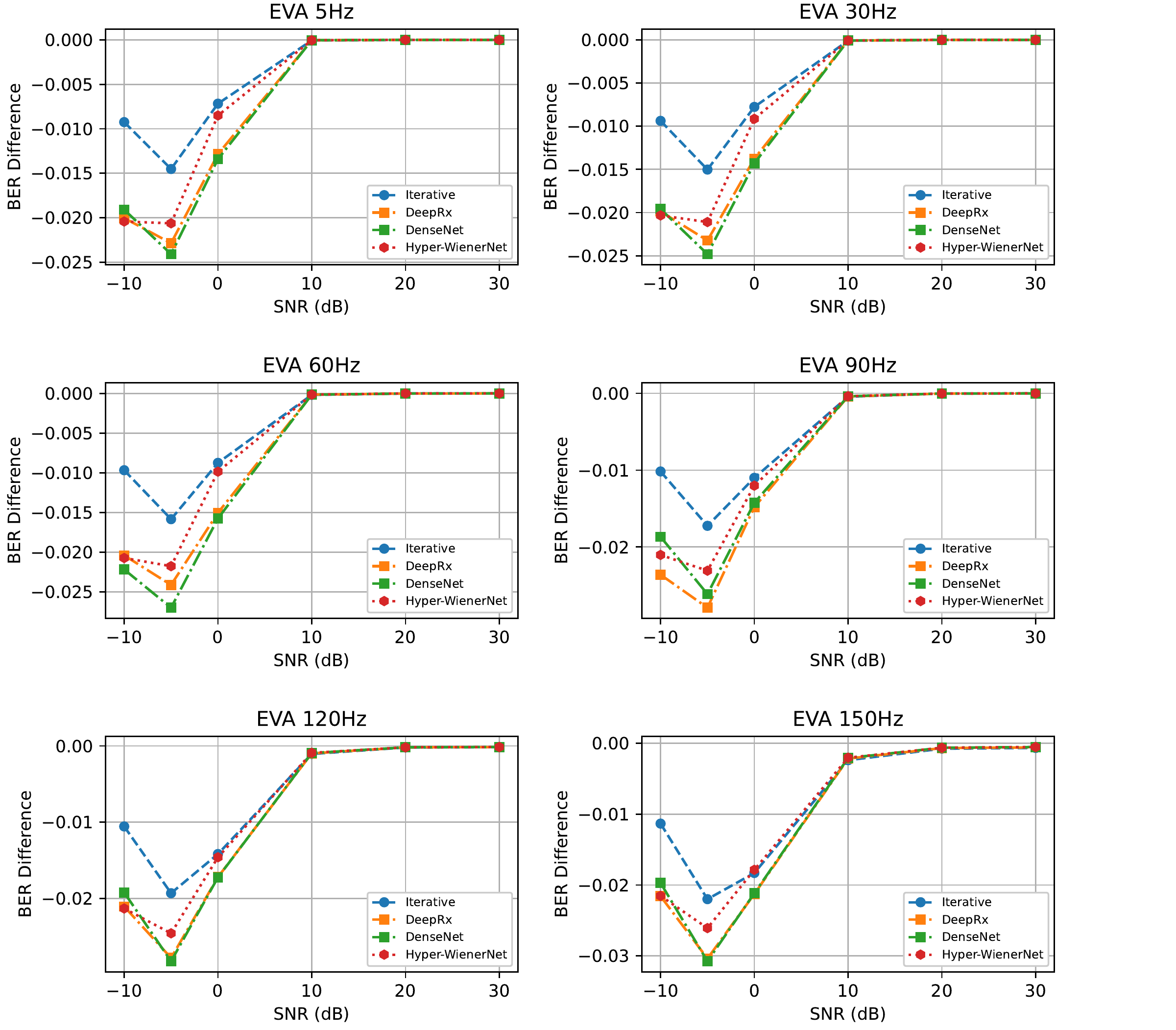}
\caption{BER diffference between the compared methods and the non-iterative (MMSE) algorithm on the EVA dataset.}
\label{fig:evadiff}
\end{figure*}

We evaluate the traditional algorithms and DL models on the test set of the EVA dataset described in Subsection \ref{subsec:datasets}. Fig. \ref{fig:eva} shows the BER performance of the compared methods on the EVA dataset. In the high-SNR scenarios ($10$ dB, $20$ dB, $30$ dB), the iterative algorithm outperforms the DL methods with a lower BER. However, in the low-SNR scenarios (-10 dB, -5 dB, 0 dB), the BER curves of the compared methods in Fig. \ref{fig:eva} are almost indistinguishable. To better discriminate between the BER curves, we plot the BER difference between the compared methods and the non-iterative algorithm in Fig. \ref{fig:evadiff}. We observe that the DL methods outperform the iterative algorithms in the low-SNR setting. Furthermore, DenseNet has better in-distribution generalization performance than Hyper-WienerNet and is comparable to DeepRx.

\begin{figure*}[!t]
\centering
\includegraphics[width=5in]{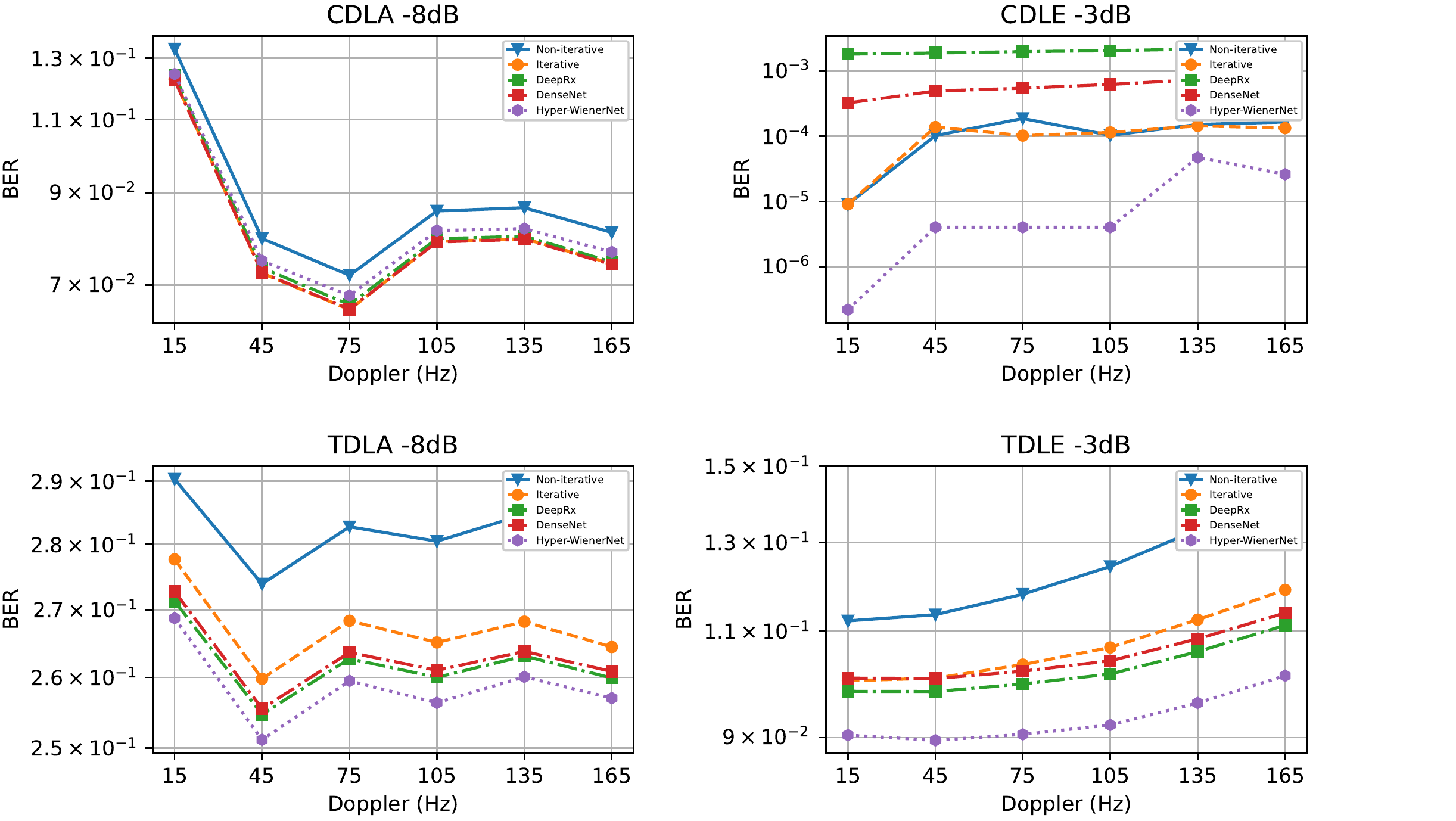}
\caption{OOD generalization results. The Doppler variable represents the maximum Doppler spread among channel paths.}
\label{fig:ood}
\end{figure*}

Fig. \ref{fig:ood} shows the OOD generalization results on channel models different the EVA channel used for training. Specifically, we evaluate the traditional algorithms and DL models on the CDL-A, CDL-E, TDL-A, TDL-E channel models. Hyper-WienerNet performs better than the other methods on most of the OOD datasets, except for CDL-A. This observation indicates that estimating data-adaptive hyperparameters via hypernetworks helps improve OOD generalization. We also note that the iterative algorithm is outperformed by DL methods in most of the low-SNR OOD datasets, which is consistent with the inferior performance of the iterative algorithm in the low-SNR setting of Fig. \ref{fig:evadiff}. Overall, these results suggest that DL methods are superior to iterative algorithms in challenging settings.

\begin{figure*}[!t]
\centering
\includegraphics[width=5in]{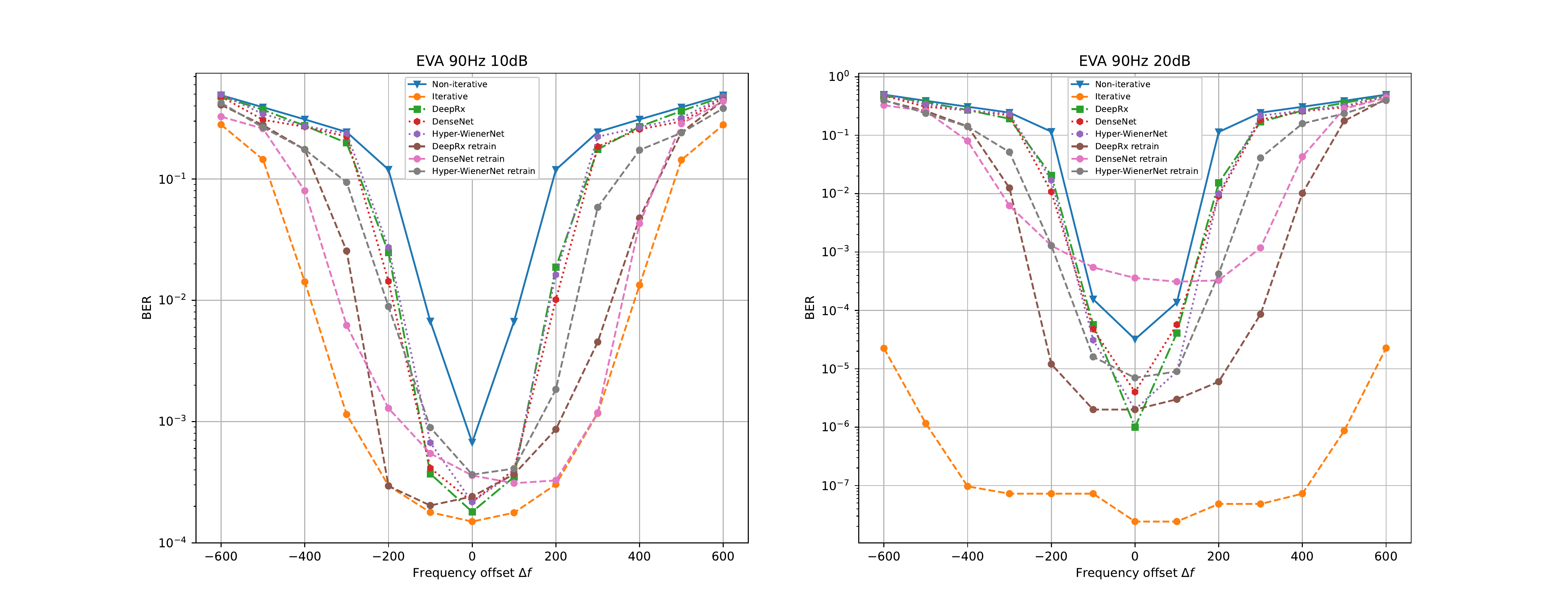}
\caption{Robustness against the carrier frequency offset.}
\label{fig:robust}
\end{figure*}

\begin{figure*}[!t]
\centering
\includegraphics[width=5in]{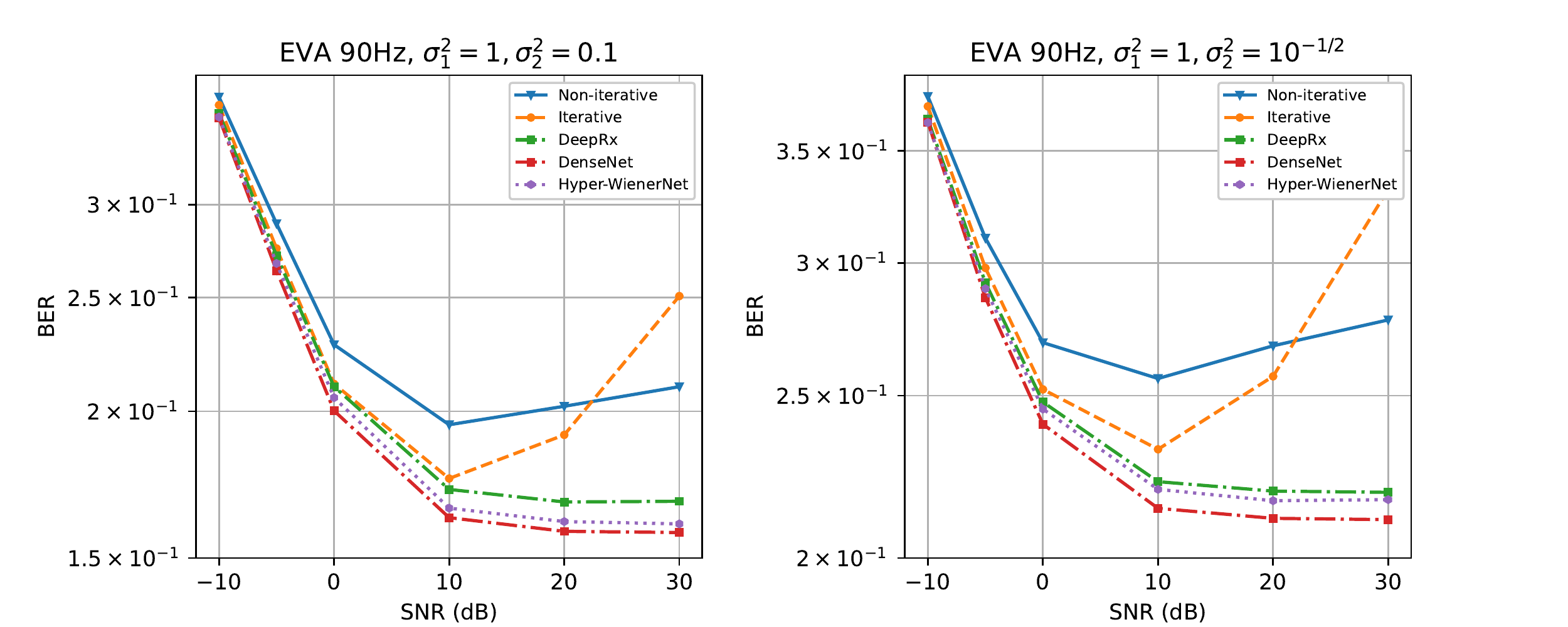}
\caption{Robustness against the asymmetric gaussian noise.}
\label{fig:noiseinterf}
\end{figure*}

\subsection{Robustness}\label{subsec:robust}
In real-world scenarios, the input data are often contaminated with noise which may degrade the model performance. In this subsection, we validate the robustness of traditional algorithms and DL models in the presence of carrier frequency offset and asymmetric Gaussian noise.

\subsubsection{Carrier Frequency Offset}
Carrier frequency offset often occurs in high-speed railway scenarios. This phenomenon causes a frequency shift in the received signal. The data affected by this shift can be mathematically represented as:
\begin{equation}
    \bm{Y}_{f,s}^{\text{shift}} = \bm{Y}_{f,s}\exp(2\pi i \Delta f\frac{2192}{15\times 10^{3}\times 2048}s),
\end{equation}
where $\Delta f$ is the offset parameter.

As shown in Fig. \ref{fig:robust}, the BER of the iterative algorithm tested on the (EVA, $90$ Hz, $10$ dB) and (EVA, $90$ Hz, $20$ dB) dataset is several orders of magnitude lower than those of DL models. Among the deep models tested on the (EVA, $90$ Hz, $10$ dB) dataset, DeepRx achieves lower BER for $|\Delta f|\leq 100$ while DenseNet is more robust around $|\Delta f|=200$. Hyper-WienerNet achieves $5\times$ lower BER than DeepRx and DenseNet on the (EVA, $90$ Hz, $20$ dB) dataset when $\Delta f=100$. In general, the DL models have similar performances in most settings.

To improve the robustness of deep models, we augment the training set with the shifted data. First, we train the model on the original dataset without the carrier frequency offset for $N_{\text{pretrain}}$ epochs. Then, the shifted data are fed into the model, where the offset parameter $\Delta f$ increases every 5 epochs. Specifically, the offset parameter $\Delta f$ for epoch $N$ ($N>N_{\text{pretrain}}$) is $\Delta f_N = 10 \lceil \frac{N-N_{\text{pretrain}}}{5}\rceil.$

Fig. \ref{fig:robust} shows the results of deep models after retraining. Our retraining strategy significantly improve the performance of deep models for relatively large offset, \textit{i.e.}, $200\leq |\Delta f|\leq 400$, but they still perform worse than the iterative algorithm. This observation indicates that this iterative algorithm is more robust to carrier frequency offsest than the compared deep models, both with and without retraining. Additionally, we observe that the performance on the original data ($\Delta f=0$) deteriorates after retraining. This phenomenon is similar to the robustness-accuracy tradeoff observed in computer vision\cite{zhang2019theoretically} and linear inverse problems\cite{colbrook2022difficulty}. Currently, it's still an open question whether it is possible to train a deep network that is both robust and accurate.

\subsubsection{Asymmetric Gaussian Noise Interference}
We add different levels of Gaussian noise to the two resource blocks, \textit{i.e.},
\begin{equation}
    \bm{Y}_{f,s,n_r}^{\text{agn}} = \bm{Y}_{f,s,n_r}+\bm{N}_{f,s,n_r},f=0,\cdots,\frac{F}{2}-1
\end{equation}
\begin{equation}
    \bm{Y}_{f,s,n_r}^{\text{agn}} = \bm{Y}_{f,s,n_r}+\bm{N}'_{f,s,n_r},f=\frac{F}{2},\cdots,F-1
\end{equation}
where $\bm{N}_{f,s,n_r}\sim\mathcal{CN}(0,2\sigma_1^2), \bm{N}'_{f,s,n_r}\sim\mathcal{CN}(0,2\sigma_2^2)$. We test the traditional algorithms and DL models in two settings: $(\sigma_1^2,\sigma_2^2)=(1,0.1)$ and $(1,10^{-1/2})$. 
Fig. \ref{fig:noiseinterf} shows the BER performance of the compared methods tested on the EVA dataset with 90 Hz Doppler and varying SNR values. DL methods exhibit superior performance compared to traditional algorithms across all SNR settings, and their advantage increases with higher SNR levels. This suggests that DL methods are more robust to asymmetric Gaussian noise interference compared to traditional algorithms.

\subsection{Complexity}
The number of floating point operations (FLOPs) and parameters of the compared methods are presented in Table \ref{tab:flops+params}. The results in Subsection \ref{subsec:quantitative}, \ref{subsec:robust} and Table \ref{tab:flops+params} show that DenseNet achieves a performance that is comparable to the previous state-of-the-art model DeepRx, while using substantially fewer parameters. However, the low-complexity iterative algorithm outperforms DL methods in high-SNR settings and is also more robust to corrupted data with carrier frequency shift. The computationally expensive deep models only perform better than the iterative algorithm in low-SNR settings. Therefore, the iterative algorithm is still a viable solution for high-SNR scenarios where low complexity and high performance are desired, while DL methods can provide improved performance in low-SNR settings.

\begin{table}[h]
\renewcommand{\arraystretch}{0.7}
\centering
\footnotesize
\begin{tabular}{c|c|c}
\toprule Method & FLOPs & \#Params     \\
\midrule Non-iterative & 340 992 &  -    \\
\midrule Iterative & 3 390 912 &  -   \\
\midrule Hyper-WienerNet & 2 451 314 880 &  4 408 718  \\
\midrule DeepRx\cite{Honkala2021DeepRxFC}  & 355 129 344 & 1 233 088  \\
\midrule DenseNet & 61 378 560 & 213 120 \\
\bottomrule
\end{tabular}
\caption{Computational cost of the compared methods.}
\label{tab:flops+params}
\end{table}

\section{Conclusion}\label{sec:conclusion}
In this paper, we establish a comprehensive benchmark for the traditional algorithms and DL methods used in JCESD and discuss the advantages and disadvantages of DL. Specifically, we evaluate three DL methods: DeepRx\cite{Honkala2021DeepRxFC}, a lightweight DenseNet adapted to JCESD, and a new UD model called Hyper-WienerNet, which uses hypernetworks to estimate unknown parameters. Our results demonstrate that Hyper-WienerNet achieves superior OOD generalization performance across various channel models, while the purely data-driven methods perform better in in-distribution generalization. Additionally, our findings indicate that the DL methods exhibit superior performance over traditional algorithms in low-SNR settings, whereas the iterative algorithm proves to be more effective in high-SNR scenarios. Furthermore, we find that DL methods are more robust to asymmetric Gaussian noise, while the iterative algorithm is considerably more robust to carrier frequency offset. Overall, our study provides insight into the strengths and limitations of traditional algorithms and DL methods for JCESD. The results suggest that there is no one-size-fits-all solution, and the choice of approach may depend on the specific application and operating conditions. Further research is needed to improve the performance of DL approaches and develop more robust and efficient algorithms for JCESD in practical wireless communication systems.



\bibliographystyle{elsarticle-num}
\bibliography{ref}

\end{document}